\documentclass[a4paper]{jpconf}
\usepackage{graphicx}
\usepackage{amssymb}

\begin{document}
\title{Kinetic Simulations of Rayleigh-Taylor Instabilities}

\author{Irina Sagert$^1$, Wolfgang Bauer$^{2,3}$, Dirk Colbry$^3$, Jim Howell$^3$, Alec Staber$^3$, Terrance Strother$^4$}
\address{$^1$ Center for Exploration of Energy and Matter, Indiana University, Bloomington, Indiana, 47308, USA\\
$^2$Department of Physics and Astronomy, Michigan State University, East Lansing, Michigan, 48824, USA\\
$^3$Institute for Cyber-Enabled Research, Michigan State University East Lansing, Michigan 48824, USA\\
$^4$XTD-6, Los Alamos National Laboratory, Los Alamos, New Mexico 87545, USA}
\ead{isagert@indiana.edu}

\begin{abstract}
We report on an ongoing project to develop a large scale Direct Simulation Monte Carlo code. The code is primarily aimed towards applications in astrophysics such as simulations of core-collapse supernovae. It has been tested on shock wave phenomena in the continuum limit and for matter out of equilibrium. In the current work we focus on the study of fluid instabilities. Like shock waves these are routinely used as test-cases for hydrodynamic codes and are discussed to play an important role in the explosion mechanism of core-collapse supernovae. As a first test we study the evolution of a single-mode Rayleigh-Taylor instability at the interface of a light and a heavy fluid in the presence of a gravitational acceleration. To suppress small-wavelength instabilities caused by the irregularity in the separation layer we use a large particle mean free path. The latter leads to the development of a diffusion layer as particles propagate from one fluid into the other. For small amplitudes, when the instability is in the linear regime, we compare its position and shape to the analytic prediction. Despite the broadening of the fluid interface we see a good agreement with the analytic solution. At later times we observe the development of a mushroom like shape caused by secondary Kelvin-Helmholtz instabilities as seen in hydrodynamic simulations and consistent with experimental observations. 
\end{abstract}

\section{Introduction}
Kinetic transport simulations are applied to study systems that have large Knudsen numbers $K > 0.01$ and therefore cannot be described by fluid dynamics \cite{Agarwal01}. Instead, they require the explicit solution of the transport equations for the particles they are composed of. The most famous example of a transport equation is the Boltzmann equation:
\begin{eqnarray}
\frac{\partial f_i (\vec{r}, \vec{p}, t) }{\partial t} + \frac{ \vec{p}}{m_i} \cdot \nabla f_i (\vec{r}, \vec{p}, t) + \vec{F} \cdot \frac{\partial f_i (\vec{r}, \vec{p}, t)}{\partial \vec{p}} = I_{i,\mathrm{ coll}} ,
\label{boltzmann} 
\end{eqnarray}
where $\vec{r}$, $\vec{p}$, and $m_i$ are the position, momentum, and mass of particle $i$ respectively, $\vec{F}$ is an external force field, and $I_{i,\mathrm{ coll}}$ takes into account the changes in the phase space distribution function $f_i (\vec{r}, \vec{p}, t)$ caused two-body collisions.\\ 
Typical examples of non-equilibrium environments are nuclei in heavy-ion collisions \cite{Bauer86, Bertsch84, Kruse85, Aichelin85, Aichelin86,Bouras12,Kortemeyer95}, hypersonic flow in aerospace research \cite{Li09}, inertial confinement fusion (ICF) capsules \cite{Casanova91,Vidal93,Vidal95}, and astrophysical systems. The latter range from flow in accretion discs \cite{Matsuda02}, cluster formation in the crust of neutron stars \cite{Schneider13}, to cosmological simulations \cite{Hernquist92}. Depending on the system, different numerical approaches are chosen (with extensions to include quantum and relativistic effects), such as Molecular Dynamics \cite{MD_primer}, Particle-In-Cell \cite{PIC}, and Direct Simulation Monte Carlo \cite{Bird94}. The latter approximates $f_i (\vec{r}, \vec{p}, t)$ by so-called test-particles in form of delta-functions:
\begin{eqnarray}
f(\vec{r},\vec{p},t)=\sum_{i=0}^{N} \delta^{3}\big(\vec{r}-\vec{r}_i(t)\big)\delta^{3}\big(\vec{p}-\vec{p}_i(t)\big) .
\end{eqnarray}
For small Knudsen numbers $K = l/L \ll 1$, when the particles' mean free path $l$ becomes smaller than a characteristic length scale $L$ of the system, kinetic simulations have been shown to reproduce hydrodynamic phenomena such as relativistic shock wave evolution \cite{Bouras09, Bouras12} and fluid instabilities \cite{Kadau10, Gan11, Ashwin10}. Therefore, transport models seem to offer the intriguing possibility to study such phenomena in equilibrium and for systems where the Navier-Stokes equations are not applicable anymore.\\ 
\newline
Our main interest lies in the study of core-collapse supernovae and inertia confinement fusion capsules. Despite the differences in both systems - ranging from scale to micro-physical processes - similar mechanisms might be at play to drive the dynamical evolution. Core-collapse supernovae are explosions of massive stars caused by the gravitational collapse of the iron core at the end of their lives \cite{Bethe90, Janka07}. Despite its rich research history \cite{Janka12}, understanding the mechanism that drives the supernova explosion remains an outstanding problem and modern computer simulations point towards a crucial role of fluid instabilities coupled to neutrino-matter interactions \cite{Janka12, Ott13, Burrows13, Sumiyoshi12, Bruenn13}. However, the modeling of both, especially in 3D and in the general relativistic regime is a major computational challenge and requires supercomputer facilities.\\ 
Inertia confinement fusion is a candidate to create fusion energy in the laboratory with current experiments being run at the National Ignition Facility \cite{Lindl95, Lindl04, Glenzer12, Edwards13}. The goal is to reach fusion via the compression and heating of fusion fuel in form of capsules containing deuterium and tritium with high-energy laser beams. Though significant progress has been made towards achieving conditions necessary for ignition, only recently a greater than unity fuel gain was reported, i.e. the energy generated through fusion reactions exceeded the amount of deposited energy into the fusion fuel \cite{Hurricane14}. Similar to core-collapse supernovae, the difficulties seem to lie in the control of fluid instabilities \cite{Bodner74} and non-equilibrium particle transport \cite{Rosenberg14} (and references therein).
\section{Rayleigh-Taylor Instability}
Motivated by the challenges of both, core-collapse supernovae and inertia confinement fusion capsules, we are developing a large scale Direct Simulation Monte Carlo code that is written to run in parallel and can model the evolution of $> 10^6$ test-particles \cite{Howell13, Sagert14, Sagert13}. The ability of the code to reproduce hydrodynamic shock wave phenomena has already been demonstrated for the continuum limit as well as for non-equilibrium matter \cite{Sagert14,Sagert13}. In this work we focus on the modeling of fluid instabilities. As a first test-case we choose the Rayleigh-Taylor instability \cite{Liska03,Jun95}. It is created when a dense fluid lies on top of a light one in the presence of a gravitational acceleration. Both fluids are initially at rest, however, perturbations of the interface become unstable, grow with time, and eventually become chaotic.\\ 
For early times the growth of an single mode perturbation $\eta_0(x)$ of wavelength $\lambda$ and amplitude $B$ can be predicted from linear theory \cite{chandra,Frieman54}: 
\begin{eqnarray}
\eta(x,t) = \frac{1}{2} \left(e^{\gamma t} + e^{- \gamma t} \right) \eta_0(x).
\label{eta}
\end{eqnarray}
Here, $\gamma = \sqrt{A g \alpha}$ is the growth rate and is given by the Atwood number $A = \frac{\rho_2 - \rho_1}{\rho_2 + \rho_1}$, the wave number $\alpha = 2 \pi / \lambda$, and the gravitational acceleration $g$. It can be seen from eq.(\ref{eta}) and $\gamma \propto \lambda^{-0.5}$ that short wavelength perturbations will grow most rapidly. The linear theory solution is valid up to a point, when the amplitude of the perturbation is of the order of $\sim 0.5 \: \lambda$. After that, the non-linear phase begins and the interface develops bubbles of light fluid entering the dense one. The latter, on the other hand, sinks in finger-shaped structures into the light fluid. This relative movement causes Kelvin-Helmholtz instabilities to develop, primarily in form of vortices at the tip of the fingers, leading to a typical mushroom shape. At later times, more secondary instabilities set in, eventually resulting in chaotic behavior. Hereby, for the same initial conditions, the evolution of secondary instabilities seem to depend on the resolution of the simulation and the applied finite difference scheme \cite{Liska03}.
\section{Simulation Setup}
In this work we study a 2D single mode Rayleigh-Taylor fluid instability applying $4 \cdot 10^7$ test-particles. Details of our code, for example the collision algorithm, can be found in \cite{Sagert14}. For the current simulation we use periodic boundary conditions and a value of $g=0.1$ for the gravitational acceleration. The width and height of the simulations box are $0 \leq x \leq 0.25$ and $0 \leq y \leq 1.6$, respectively, whereas we divide the simulation space into $800 \times 5120$ bins for the calculation and $200 \times 640$ bins for the output. Hereby, the upper half of the box with $y \geq 0.8$ (2) is filled with the high density gas while the lower half (1) contains low density matter with $\rho_2 = 2 \rho_1$. The volumes $V_2$ and $V_1$ are the same and, to save computational time, we fill them with the same number of test-particles $N_2 = N_1$. Consequently, the latter are given different masses $m_2 = 2 \: m_1$. Furthermore, to better follow the evolution of the instability, particles in the upper half are assigned a type-value $\tau =2$ while particles in the lower half have $\tau = 1$. To initialize the single mode instability we perturb the separation layer between fluid (2) and (1) according to: 
\begin{eqnarray}
\eta(x) = 0.8 + B \cos(2 \pi x/\lambda),
\end{eqnarray}
where we set the perturbation amplitude to $B=0.01$ and its wavelength to $\lambda = 0.5$. A particle absolute velocity $v_i$ is then determined according to the Maxwell-Boltzmann distributions with a height dependent temperature that is given by the barometric formula and the particle's $y$-position $y_i$:
\begin{eqnarray}
T (h_i) = (V_{1,2}/N_{1,2}) \left( P_0 - \rho_{1,2} \: g \: h_i \right) , \: \: h_i = y_i - 0.8, 
\end{eqnarray}
whereas we choose $P_0 = 2.5$.\\ 
For the initial distribution of particle velocities we use a Monte Carlo algorithm that works in the following way: From the height $h_i$ we determine the temperature and root-mean-square velocity of particle $i$: 
\begin{eqnarray}
v_{\mathrm{rms,i}} = \sqrt{2\:T(h_i)/m_i}.
\end{eqnarray}
We assign a velocity $v_i \in [0:6 \: v_{\mathrm{rms}}]$ and determine its probability according to the normalized Maxwell-Boltzmann velocity distribution:  
\begin{eqnarray}
\mathcal{P} (v_i) = \sqrt{\frac{m_i \: e}{kT}} \: v_i \: \exp{ \left(- \frac{m_i}{2\: kT} v_i^2 \right)}.
\end{eqnarray}
Then, we choose a random probability $\mathcal{P}_r \in [0:1]$. If $\mathcal{P}_r  \leq \mathcal{P} (v_i)$, $v_i$ is assigned to particle $i$, otherwise we repeat the procedure.\\
Unlike hydrodynamics codes, a particle-based approach always introduces some unevenness of the fluid interface due to the finite number of particles. This can serve as a potential seed for small-wavelength instabilities that, as mentioned previously, have the fastest growth rate and could thereby impact the evolution of the single mode perturbation. In order to prevent the development of these instabilities we set the particle mean free path to $l = 2.0 \: dx$, where $dx$ is the width of a bin. With that, we decrease the resolution of our simulation and blur possible irregularities of the interface. But, it should also be noted that a large mean free path increases the amount of particles diffusing from one fluid into the other. This leads to the development of a diffusion layer that could also influence the evolution of the single-mode instability.  
\section{Results}
\begin{figure}
\begin{center}
\includegraphics[width = 16cm]{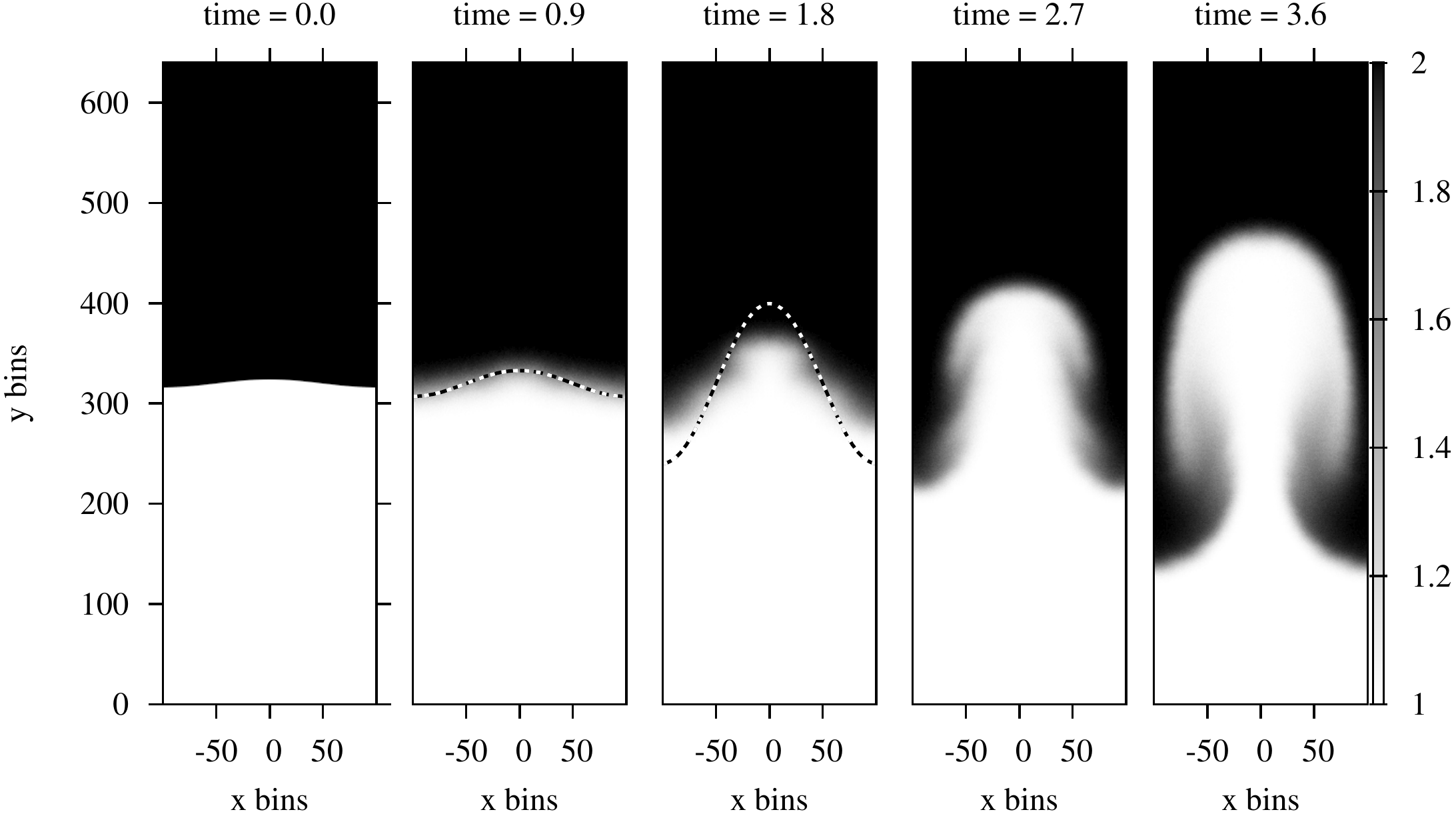}
\caption{Average particle type $\tau$ in the Rayleigh-Taylor instability with $4 \cdot 10^7$ test particles and $\lambda = 2 \: d x$ at different simulation times together with the analytic prediction for the fluid interface taken from eq.(\ref{eta}). The simulation space is mirrored at $x=0$.}
\label{figure}
\end{center}
\end{figure}

\begin{figure}
\begin{center}
\includegraphics[width = 16cm]{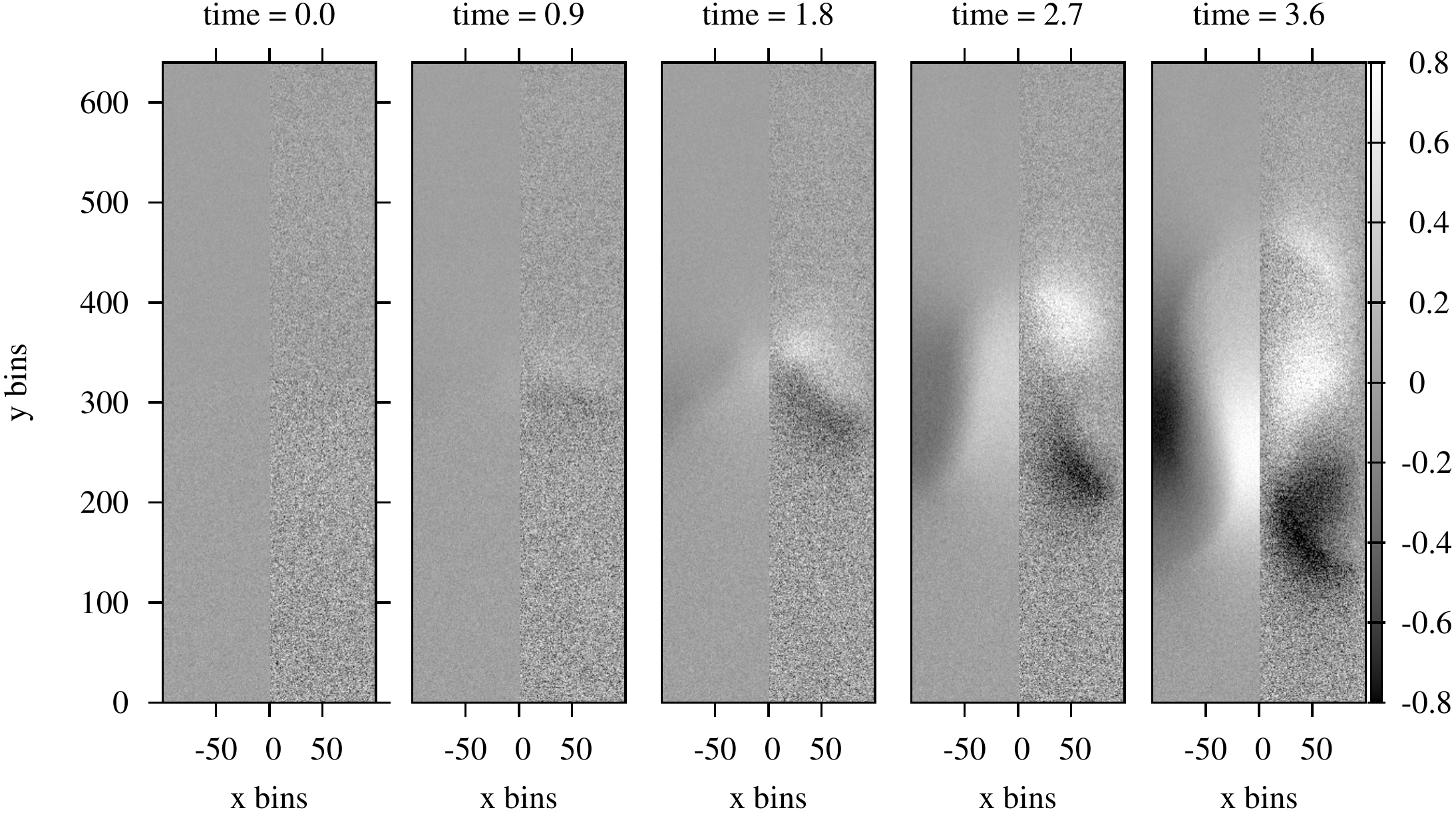}
\caption{Average $x$ and $y$ particle velocities per bin at different simulation times. $y$-velocities are shown for $x \leq 0$, while $x$-velocities are shown for $x > 0$ and are scaled by a factor of 4 for better comparison.}
\label{figure2}
\end{center}
\end{figure}

Figure \ref{figure} shows the average particle type $\tau$ per bin in our 2D Rayleigh-Taylor simulation at different times together with the analytic prediction for $\eta(x,t)$. The number of time steps to reach a simulation time of $t = 3.6$ is of the order of $10^5$ whereas the size of the time steps is adaptive and determined from the maximal particle velocity \cite{Sagert14}.
For $t \lesssim 0.5$ we observe a broadening of the interface as particles diffuse from one fluid into the other. Despite the latter and the difficulty to locate its precise position, the layer seems to stay within the analytic prediction. For times $t \gtrsim 1.5$, the simulation enters the non-linear regime as the amplitude of the perturbation becomes comparable to $\sim 0.5 \: \lambda$. Although we blur the fluid interface via the large particle mean free path, small structures are developing and are visible starting at $t \sim 1.1$. At later times, we observe the formation of the low density bubble and its rise in into the high-density matter whereas the latter sinks at the edges of the simulation box into the light fluid. Figure \ref{figure2} shows the average $x$ and $y$ particle velocities per bin for the same simulation times as in Figure \ref{figure}. Here, the $y$-velocity is plotted in the left half of the simulation boxes (i.e. for $x<0$) while the $x$-velocity is plotted on the right hand side. For better comparison we scale the $x$-velocity by a factor of 4. It can be seen that a weak vorticity seems to develop leading to the expected mushroom shape of the instability. However, the dynamics is dominated by the motion into the positive and negative y-directions.\\
Overall, we observe the appearance of the typical Rayleigh-Taylor structure on expected time scales.  
Differences to hydrodynamic simulations lie in the growth rate which seems to overestimate the position of the instability in our simulation for $t \gtrsim 1.5$ as well as the strength of secondary Kelvin-Helmholtz instabilities. It is interesting to note that hydrodynamic simulations find a decrease in the growth rate of the single mode Rayleigh-Taylor instability and a delayed appearance of Kelvin-Helmholtz instabilities for low Reynolds numbers $\mathrm{Re}_p = (\lambda/ \nu) \sqrt{A g \lambda /(1+A)}$ where $\nu$ is the viscosity \cite{Wei12}. More kinetic studies should be conducted to examine any possible growth rate dependence on quantities such as test-partcle mean free path or the number of test-particles used to represent the fluids being modeled \cite{Sagert_prep}. 
\section{Summary and Outlook}
In this work we present the first study of a single mode Rayleigh-Taylor instability performed with our large-scale Direct Simulation Monte Carlo code. The simulation uses $4 \cdot 10^7$ test-particles and follows the evolution of the instability from the linear to the non-linear regime for a simulation time of $t=3.6$ in $\sim 10^5$ time steps. To prevent the early formation of small perturbations caused by irregularities of the interface between the light and heavy fluids we set the particle mean free path to a value of $l = 2 \: dx$. With that, small scale irregularities are blurred, however, particles are also able to propagate from one fluid into the other leading to the formation of a diffusion layer. For simulation times $t \lesssim 1.5$ we find that the position of the interface generally agrees with the analytic prediction from linear theory. For later times a growing overestimate of the latter can be seen as the amplitude of the simulated instability lags behind the analytic solution. A possible explanation can be that at these times, linear theory is not valid anymore as the instability enters the non-linear regime. Weak secondary Kelvin-Helmholtz instabilities develop at later times and result in the formation of the typical mushroom shape. However, the effect is weak as the dynamics are dominated by the motion in the positive and negative $y$ directions. We find that the observed evolution is similar to hydrodynamic simulations with low Reynolds numbers. Future studies will include a more detailed analysis of the Rayleigh-Taylor instability applying different values for the particle mean free path as well as the total number of test-particles. 
\section{Acknowledgements}
This work used the Extreme Science and Engineering Discovery Environment (XSEDE), which is supported by National Science Foundation grant number OCI-1053575. Furthermore, I.S. acknowledges the support of the High Performance Computer Center and the Institute for Cyber-Enabled Research at Michigan State University. 
\section*{References}
\bibliographystyle{iopart-num}
\bibliography{ref}

\providecommand{\newblock}{}
\begin{thebibliography}{10}
\expandafter\ifx\csname url\endcsname\relax
  \def\url#1{{\tt #1}}\fi
\expandafter\ifx\csname urlprefix\endcsname\relax\def\urlprefix{URL }\fi
\providecommand{\eprint}[2][]{\url{#2}}

\bibitem{Agarwal01}
{Agarwal} R~K, {Yun} K~Y and {Balakrishnan} R 2001 {\em Physics of Fluids\/}
  {\bf 13} 3061--3085

\bibitem{Bauer86}
{Bauer} W, {Bertsch} G~F, {Cassing} W and {Mosel} U 1986 {\em Physical Review
  C\/} {\bf 34} 2127--2133

\bibitem{Bertsch84}
{Bertsch} G~F, {Kruse} H and {Gupta} S~D 1984 {\em Physical Review C\/} {\bf
  29} 673--675

\bibitem{Kruse85}
{Kruse} H, {Jacak} B~V and {St{\"o}cker} H 1985 {\em Physical Review Letters\/}
  {\bf 54} 289--292

\bibitem{Aichelin85}
{Aichelin} J and {Bertsch} G 1985 {\em Physical Review C\/} {\bf 31} 1730--1738

\bibitem{Aichelin86}
{Aichelin} J and {St{\"o}cker} H 1986 {\em Physics Letters B\/} {\bf 176}
  14--19

\bibitem{Bouras12}
{Bouras} I, {El} A, {Fochler} O, {Niemi} H, {Xu} Z and {Greiner} C 2012 {\em
  Physics Letters B\/} {\bf 710} 641--646

\bibitem{Kortemeyer95}
{Kortemeyer} G, {Bauer} W, {Haglin} K, {Murray} J and {Pratt} S 1995 {\em
  Physical Review C\/} {\bf 52} 2714--2724

\bibitem{Li09}
{Li} Z~H and {Zhang} H~X 2009 {\em Journal of Computational Physics\/} {\bf
  228} 1116--1138

\bibitem{Casanova91}
Casanova M, Larroche O and Matte J~P 1991 {\em Physical Review Letter\/} {\bf
  67}(16) 2143--2146

\bibitem{Vidal93}
Vidal F, Matte J~P, Casanova M and Larroche O 1993 {\em Physics of Fluids B:
  Plasma Physics\/} {\bf 5} 3182--3190

\bibitem{Vidal95}
Vidal F, Matte J~P, Casanova M and Larroche O 1995 {\em Physical Review E\/}
  {\bf 52}(4) 4568--4571

\bibitem{Matsuda02}
{Matsuda} T, {Mizutani} H and {Boffin} H~M~J 2002 {\em The Physics of
  Cataclysmic Variables and Related Objects\/} ({\em Astronomical Society of
  the Pacific Conference Series\/} vol 261) ed {G{\"a}nsicke} B~T, {Beuermann}
  K and {Reinsch} K p 505

\bibitem{Schneider13}
{Schneider} A~S, {Horowitz} C~J, {Hughto} J and {Berry} D~K 2013 {\em Physical
  Review C\/} {\bf 88} 065807

\bibitem{Hernquist92}
{Hernquist} L and {Ostriker} J~P 1992 {\em The Astrophysical Journal\/} {\bf
  386} 375--397

\bibitem{MD_primer}
Ercolessi F 1997 A molecular dynamics primer Spring College in Computational
  Physics, ICTP, Trieste

\bibitem{PIC}
Tskhakaya D 2008 {\em Computational Many-Particle Physics\/} ({\em Lecture
  Notes in Physics\/} vol 739) ed Fehske H, Schneider R and Wei§e A (Springer
  Berlin Heidelberg) pp 161--189 ISBN 978-3-540-74685-0

\bibitem{Bird94}
Bird G~A 1994 {\em Molecular gas dynamics and the direct simulation of gas
  flows\/} (Clarendon Press)

\bibitem{Bouras09}
{Bouras} I, {Moln{\'a}r} E, {Niemi} H, {Xu} Z, {El} A, {Fochler} O, {Greiner} C
  and {Rischke} D~H 2009 {\em Physical Review Letters\/} {\bf 103} 032301

\bibitem{Kadau10}
{Kadau} K, {Barber} J~L, {Germann} T~C, {Holian} B~L and {Alder} B~J 2010 {\em
  Royal Society of London Philosophical Transactions Series A\/} {\bf 368}
  1547--1560

\bibitem{Gan11}
{Gan} Y, {Xu} A, {Zhang} G and {Li} Y 2011 {\em Physical Review E\/} {\bf 83}
  056704

\bibitem{Ashwin10}
{Ashwin} J and {Ganesh} R 2010 {\em Physical Review Letters\/} {\bf 104} 215003

\bibitem{Bethe90}
{Bethe} H~A 1990 {\em Reviews of Modern Physics\/} {\bf 62} 801--866

\bibitem{Janka07}
{Janka} H~T, {Langanke} K, {Marek} A, {Mart{\'{\i}}nez-Pinedo} G and
  {M{\"u}ller} B 2007 {\em Physics Reports\/} {\bf 442} 38--74

\bibitem{Janka12}
{Janka} H~T 2012 {\em Annual Review of Nuclear and Particle Science\/} {\bf 62}
  407--451

\bibitem{Ott13}
{Ott} C~D, {Abdikamalov} E, {M{\"o}sta} P, {Haas} R, {Drasco} S, {O'Connor}
  E~P, {Reisswig} C, {Meakin} C~A and {Schnetter} E 2013 {\em The Astrophysical
  Journal\/} {\bf 768} 115

\bibitem{Burrows13}
{Burrows} A 2013 {\em Reviews of Modern Physics\/} {\bf 85} 245--261

\bibitem{Sumiyoshi12}
{Sumiyoshi} K and {Yamada} S 2012 {\em The Astrophysical Journal Supplement\/}
  {\bf 199} 17

\bibitem{Bruenn13}
{Bruenn} S~W, {Mezzacappa} A, {Hix} W~R, {Lentz} E~J, {Bronson Messer} O~E,
  {Lingerfelt} E~J, {Blondin} J~M, {Endeve} E, {Marronetti} P and {Yakunin} K~N
  2013 {\em The Astrophysical Journal Letters\/} {\bf 767} L6

\bibitem{Lindl95}
{Lindl} J 1995 {\em Physics of Plasmas\/} {\bf 2} 3933--4024

\bibitem{Lindl04}
Lindl J~D, Amendt P, Berger R~L, Glendinning S~G, Glenzer S~H, Haan S~W,
  Kauffman R~L, Landen O~L and Suter L~J 2004 {\em Physics of Plasmas\/} {\bf
  11} 339--491

\bibitem{Glenzer12}
{Glenzer} S~H, {Callahan} D~A, {MacKinnon} A~J, {Kline} J~L, {Grim} G, {Alger}
  E~T, {Berger} R~L, {Bernstein} L~A, {Betti} R and {Bleuel et al} 2012 {\em
  Physics of Plasmas\/} {\bf 19} 056318

\bibitem{Edwards13}
Edwards M~J, Patel P~K, Lindl J~D, Atherton L~J, Glenzer S~H, Haan S~W,
  Kilkenny J~D, Landen O~L, Moses E~I, Nikroo A and et~al 2013 {\em Physics of
  Plasmas (1994-present)\/} {\bf 20} 070501

\bibitem{Hurricane14}
{Hurricane} O~A, {Callahan} D~A, {Casey} D~T, {Celliers} P~M, {Cerjan} C,
  {Dewald} E~L, {Dittrich} T~R, {D{\"o}ppner} T, {Hinkel} D~E, {Hopkins} L~F~B
  and {et al} 2014 {\em Nature\/} {\bf 506} 343--348

\bibitem{Bodner74}
{Bodner} S~E 1974 {\em Physical Review Letters\/} {\bf 33} 761--764

\bibitem{Rosenberg14}
{Rosenberg} M~J, {Rinderknecht} H~G, {Hoffman} N~M, {Amendt} P~A, {Atzeni} S,
  {Zylstra} A~B, {Li} C~K, {S{\'e}guin} F~H, {Sio} H, {Johnson} M~G and {et al}
  2014 {\em Physical Review Letters\/} {\bf 112} 185001

\bibitem{Howell13}
{Howell} J, {Bauer} W, {Colbry} D, {Pickett} R, {Staber} A, {Sagert} I and
  {Strother} T 2014 {\em Proceedings of Nuclear Physics: Presence and Future\/}
  ({\em FIAS Interdisciplinary Science Series, Springer Verlag\/} vol~2)

\bibitem{Sagert14}
{Sagert} I, {Bauer} W, {Colbry} D, {Howell} J, {Pickett} R, {Staber} A and
  {Strother} T 2014 {\em Journal of Computational Physics\/} {\bf 266} 191--213

\bibitem{Sagert13}
{Sagert} I, {Bauer} W, {Colbry} D, {Pickett} R and {Strother} T 2013 {\em
  Journal of Physics Conference Series\/} {\bf 458} 012031

\bibitem{Liska03}
Liska R and Wendroff B 2003 {\em Hyperbolic Problems: Theory, Numerics,
  Applications\/} ed Hou T and Tadmor E (Springer Berlin Heidelberg) pp
  831--840 ISBN 978-3-642-62929-7

\bibitem{Jun95}
{Jun} B~I, {Norman} M~L and {Stone} J~M 1995 {\em The Astrophysical Journal\/}
  {\bf 453} 332

\bibitem{chandra}
Chandrasekhar S 1961 {\em Hydrodynamic and Hydromagnetic Instabilities\/}
  (Oxford: Clarendon Press)

\bibitem{Frieman54}
{Frieman} E~A 1954 {\em The Astrophysical Journal\/} {\bf 120} 18

\bibitem{Wei12}
Wei T and Livescu D 2012 {\em Phys. Rev. E\/} {\bf 86}(4) 046405

\bibitem{Sagert_prep}
{Sagert} I and {et al} 2014 in preparation

\end{thebibliography}

\end{document}